\documentclass[preprint,showpacs,preprintnumbers,amsmath,amssymb,aps,pre]{revtex4-1}
\usepackage{graphicx}
\usepackage{dcolumn}
\usepackage{bm}
\begin{document}
\preprint{LA-UR-17-26458}
\title{Temperature dependence of dynamic slowing down \\
 in monatomic liquids from V-T theory}
\author{Duane C. Wallace}
\affiliation{Theoretical Division, Los Alamos National Laboratory, 
Los Alamos, New Mexico 87545}
\author{Giulia De~Lorenzi-Venneri}
\affiliation{Theoretical Division, Los Alamos National Laboratory, 
Los Alamos, New Mexico 87545}

\date{\today}
\begin{abstract}
For an MD  system representing a monatomic liquid, the distribution of $3N$-dimensional potential energy structures consists of two classes, random and symmetric. This distribution is shown and discussed for liquid Na. The random class constitutes the liquid phase domain. In V-T theory, the liquid atomic motion consists of prescribed vibrations in a random valley, plus parameterized transit motions between valleys. The theory has been strongly verified at 395.1K, a bit above melting. Our goal here is to test this theory for its ability to explain the temperature ($T$) dependence of the mean square displacement (MSD) at $T\leq395.1$K. The test results are positive at 204.6K, where the time evolution equations, controlled by a transit rate decreasing with $T$, accurately account  for MD data for the MSD. To test at significantly lower $T$, where the MD system does not remain in the liquid phase, the theoretical liquid MSD is calibrated for $T\leq$ 204.6K. The Kob-Andersen (K-A) dynamic slowing down graph is shown for liquid Na at $T\leq395.1$K. The following observations are discussed in terms of the atomic motion. (a) The agreement between pure vibrational motion and MD data for time correlation functions in the vibrational interval is so far highly accurate. (b) The ``bump" ahead of the plateau in the MSD at low $T$ is attributed to the vibrational excess. (c) The K-A graph from theory for liquid Na, and the same graphs from MD data for a liquid Lennard-Jones binary mixture (BMLJ)  and liquid silica, are identical in the ballistic period and in the purely diffusive time interval. (d) The glass transition proceeds in the symmetric manifold. These and other discussions confirm that V-T theory can explain the $T$ dependence observed in K-A graphs.
\end{abstract}

\pacs{05.20.Jj, 65.2Jk}
\keywords {Liquid Dynamics, diffusion, mean square displacement, V-T Theory, dynamic slowing down}
\maketitle

\section{Introduction}

Our overall goal is to develop a theory of liquid dynamics following the established techniques of condensed-matter many body theory \cite{Pines_1997_MBP, Pines_1997_EES,Kittel1963,Harrison, Ashcroft_Mermin, Glyde_book,DWbook2}. We study equilibrium thermodynamic functions and time correlation functions, both within the same vibration-transit (V-T) formulation of the atomic motion. The motion itself consists of ab initio many-atom vibrational motion plus a parameterized transit model. V-T theory for the mean square displacement, $X_{VT}(t)$, and for the self-intermediate scattering function, $F^{s}_{VT}(q,t)$, have been tested by comparison with the corresponding MD functions, $X_{MD}(t)$ and $F^{s}_{MD}(q,t)$, for liquid Na. It was recently shown that calibration of the atomic motion from $X_{MD}(t)$ produces a highly accurate theory for both $X_{VT}(t)$ and $F^{s}_{VT}(q,t)$ for all $q$  and $t$ \cite{MSD1,SISF2}.  This means we now have an explicit relation between  $X_{VT}(t)$ and $F^{s}_{VT}(q,t)$, for one liquid at one temperature. This relation is expected to be valid for a wide range of elemental liquids.

Our goal here is to test the ability of  theory to explain the dependence on temperature of the mean square displacement. To this end we shall study the theoretical $X_{VT}(t)$ by comparison with  $X_{MD}(t)$, to $T$ as low as the MD system remains an equilibrium liquid. Then we shall study the theory alone at still lower $T$. 
That theory will exhibit the same dynamic slowing down behavior as do commonly studied glass forming liquids.

In section II, a quantitative description of the entire condensed-matter potential energy surface is made 
in terms of two classes of intersecting 3$N$-dimensional valleys. In accord with general many-body theory, the random class is assigned to the liquid phase. 
The change in relative stability of random and symmetric states causes the liquid to decay into lower-lying symmetric states upon cooling.

In section III, we start with the  $X_{VT}(t)$  formulation at 395.1K, from \cite{MSD1}, and follow the same procedure to calibrate $X_{VT}(t)$ at 204.6K. 
The large relative change in $T$, from 395.1K to 204.6K, will challenge our goal. The final parameter calibrations are rationalized in sections III-IV

In section IV, we investigate one more physical behavior of low temperature liquids, dynamic slowing down. A highly informative description of this process is provided by a ``K-A (Kob-Andersen)" graph, such as the $X_{MD}(t)$ graph in figure 2 of \cite{KA1995a} for a supercooled liquid BMLJ system. We use the information from our two calibrated $X_{VT}(t)$ functions, at 
395.1K and  204.6K, to create a purely theoretical formulation at the still-lower temperatures of slow dynamics. The slowing down behavior of liquid Na is compared with that of the BMLJ and silica liquids.

In section V, the current status of V-T theory for the mean square displacement is discussed. The theory consists of three time-evolution equations, in three successive time intervals, in which characteristic combinations of vibrational and transit motions are measured by $X_{VT}(t)$. The discussion includes a recounting of valuable analytic properties of vibrational theory, and of its remarkable ability to accurately copy $X_{MD}(t)$ in the vibrational interval. In the crossover interval, vibrational motion unmasked by vanishing transit motion produces a subtle feature in $X_{VT}(t)$. Finally, dynamic slowing down in the transit random walk interval shows the same pure diffusive behavior studied long ago in MD calculations for glass forming liquids.

\section{Statistical mechanics for the supercooled liquid}

In the theoretical treatment of the atomic motion in a condensed matter system, the foundational step is to specify the 3$N$-dimensional potential surface that drives the motion. For a crystalline system, the motion is vibrations about the crystal structure, and each specific crystal structure defines a separate condensed-matter phase. No less than the crystal, the liquid also has its separate potential surface, which drives the uniquely-liquid atomic motion. In order to reveal all the physical attributes of the liquid potential surface, we shall lay out the entire condensed matter potential surface. The description was initially presented as an hypothesis, as follows \cite{DCW1997}. The potential surface is composed of intersecting 3$N$-dimensional potential  valleys in two classes, random valleys having maximal structural disorder, and symmetric valleys, having remnants of crystalline symmetry. The random valleys are of overwhelming numerical superiority, they lie highest, and are narrowly distributed.  The symmetric valleys are far fewer in number, lower in potential energy, and broadly distributed in their macroscopic statistical mechanical properties.  The hypothesis is verified by a long-running analysis of experimental data for thermodynamic properties of elemental liquids, sketched in section 1 of \cite{SISF2}. We shall now show by direct calculation that the potential surface of liquid and solid phases of Na adheres to this hypothesis.

\begin{figure} [h]
\includegraphics[height=5.0in,width=5.0in]{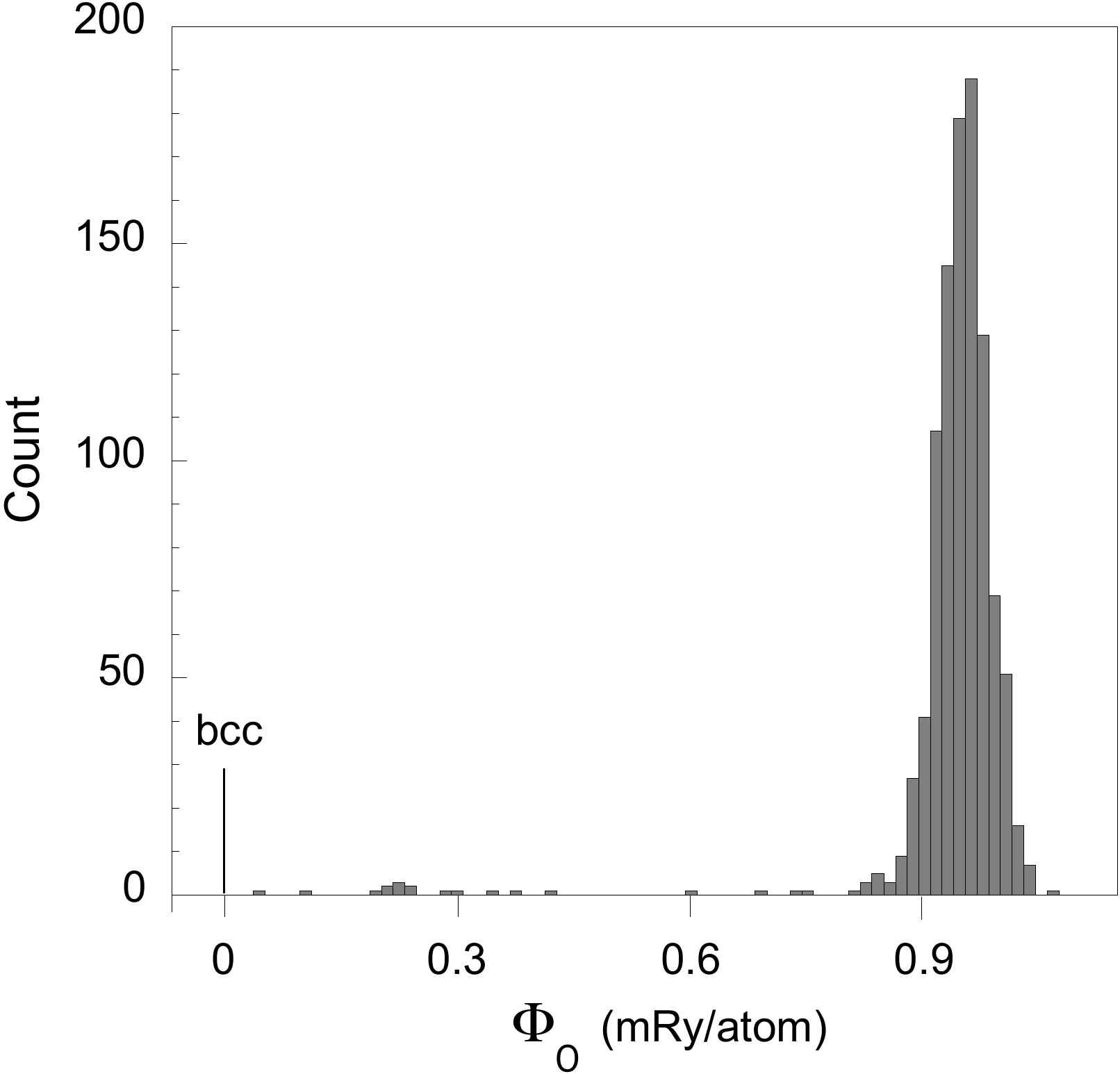}
\caption{Distribution of structure potentials $\Phi_{0}$ from 1000 quenches from initial stochastic configurations for Na at $N=500$. $\Phi_{0}$ are measured from the bcc crystal, which is calculated directly. The quenching procedure follows figure~5 of~\cite{H1}.}
\label{fig1}
\end{figure}

For the present Na system, 1000 quenches from initial stochastic configurations produced the distribution of structural potentials $\Phi_{0}$ shown in figure~\ref{fig1}. These structures have unique physical roles. Let us use superscripts $bcc$, $s$ and $r$ to identify respectively the bcc structure, or a symmetric or random structure. $\Phi_{0}^{bcc}$ is the lowest lying structure potential available to the system, and constitutes a limit point of the set $\{ \Phi_{0}^{s} \}$. The bcc crystal is the stable phase for $0\leqslant T \leqslant T_{m}$. The symmetric structures belong to higher lying crystals, amorphous solids and glasses, all metastable. As shown in figure~\ref{fig1}, the symmetrics are relatively few in number and their $\Phi_{0}^{s}$ values are broadly distributed between $\Phi_{0}^{bcc}$ and the narrow but dominant $\Phi_{0}^{r}$ distribution. That distribution, solely and entirely, belongs to the liquid phase  when the liquid is stable, at $T\geq T_{m}$. The same distribution therefore belongs to the liquid phase at all $T$. A key property of the random distribution is that its width approaches zero as $N$ increases, as shown in figures 1 and 3 of \cite{H2}. This implies that the random valleys approach macroscopic uniformity as $N$ increases, hence a single random valley is sufficient for calculating liquid statistical mechanical properties.

In an early study of Na, our MD system was cooled from above $T_{m}$, stopping at intervals to calculate equilibrium properties at a fixed $T$ \cite{WC1999}. The mean potential $\left< \Phi\right>$  decreases with $T$, until around 0.5$T_{m}$, where $\left< \Phi\right>$  spontaneously moves down from the liquid into the symmetric manifold and stays there. We also carried out quench-and-run calculations where the system approaches equilibrium in a liquid state at $T$ below $0.5 T_{m}$, but passes equilibrium and falls into the symmetric manifold. These behaviors  tell us that, among the collection of metastable states, the liquid is relatively stable above roughly $ 0.5 T_{m}$, and the symmetric manifold is relatively stable below that temperature. This change in relative stability results from the higher entropy of the liquid state, and the lower energy of the symmetric manifold, as figure \ref{fig1} shows. Since this entropy-energy competition is always present, the liquid-to-symmetric fall-down is a general condensed matter property.

The liquid-to-symmetric  transition has been studied extensively in connection with the glass transition, for monatomic and binary LJ systems \cite{JC,FA,IB,JK,IA,FC,FF,FB}. The transition depends on the cooling rate, hence is a nonequilibrium process \cite{JC,IB}. It is considered the beginning of the glass transition \cite{JC,JK}. This transition is of interest here because it functions as the low-$T$  boundary of the relatively stable equilibrium liquid.  A further comment on this topic appears in section V.C.

We close here with a practical note. In order to verify that an MD system is in an equilibrium liquid state, it must be shown that the trajectory satisfies the usual equilibrium conditions, and that the system moves among random valleys only. Because the MD system has finite $N$, one must recognize the possibility of symmetric contamination. We recall the observation of Shah and Chakravarty \cite{IE}, ``\dots  structures always contain a few configurations with a fair degree of local icosahedral or fcc/hcp order".

\section{Mean square displacement}

Our system employs a pair potential representing liquid Na at the density of the liquid at melt. The potential is well tested over many years \cite{DWbook2}, and is graphed in figure 1 of \cite{WC1999}. The mean square displacement theory is developed in \cite{MSD1}; the formulation is identical here. Comparison of the figures of this section with those in \cite{MSD1} is informative.

\begin{figure} [h]
\includegraphics[height=5.0in,width=5.0in,keepaspectratio]{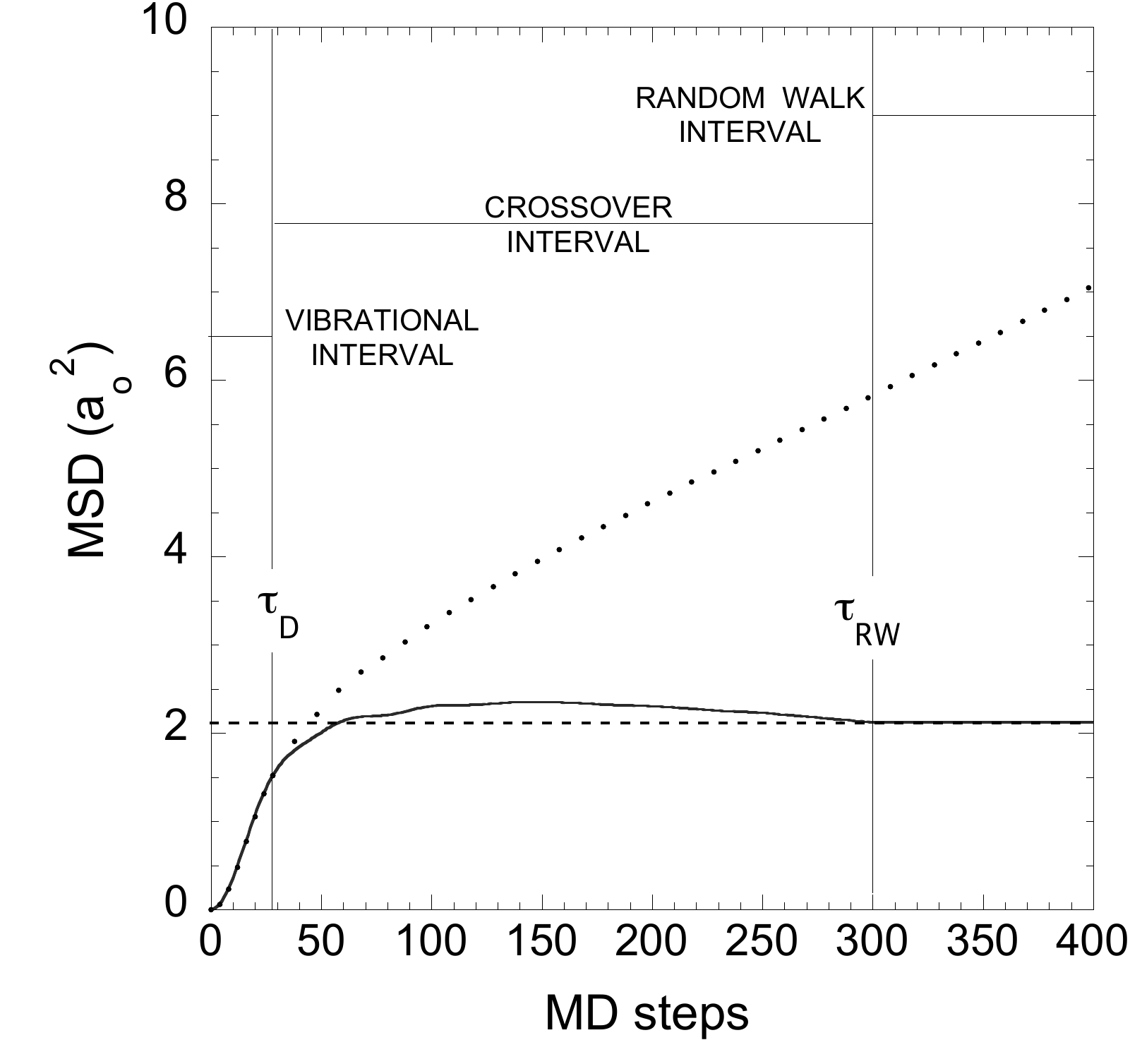}
\caption{At 204.6K: Dots are $X_{MD}(t)$, solid line is $X_{vib}(t)$, and dashed line is $X_{vib}(\infty)$. At this temperature we have $\tau_{RW}=t_{\infty}=300\delta t$ (see text). Dots are at every 4$\delta t$ to $\tau_{D}$, then every 10$\delta t$. }
\label{fig2}
\end{figure}

The atomic motion is composed of vibrational  and transit components, producing the respective contributions $X_{vib}(t)$ and  $X_{tr}(t)$, so that
\begin{equation} \label{eq1}
X_{VT}(t) = X_{vib}(t) + X_{tr}(t). 
\end{equation}
In \cite{MSD1}, we added a small interaction $X_{int}(t)$ in equation~(\ref{eq1}); this notation is suppressed here, but the interactions are fully accounted for. Figure \ref{fig2} shows $X_{MD}(t)$ and  $X_{vib}(t)$ on three time intervals, each corresponding to specific atomic-motional contributions to a time correlation function. 

The vibrational contribution is
\begin{equation} \label{eq2}
X_{vib}(t)=\frac{6 k_{B}T}{M} \frac{1}{3N-3}\sum_{\lambda} \frac{1-\cos \omega_{\lambda}t}{\omega_{\lambda}^2},
\end{equation}
where $M$ is the atomic mass, and $\omega_{\lambda}$ are the frequencies of the vibrational normal  modes $\lambda = 1, \dots, 3N-3$, having omitted the three modes for which $\omega_{\lambda}=0$. The normal mode frequencies and eigenvectors are calculated from the dynamical matrix at a representative random structure, and functions like equation~\ref{eq2} are derived from statistical mechanics theory. For time correlation functions in general, the ballistic contribution is automatically contained as the leading time dependence of the vibrational contribution. After the ballistic motion, for $t>9 \delta t$, vibrational dephasing takes over and brings the set of terms in $\cos \omega_{\lambda}t$ to zero, at the time denoted $t_{\infty}$. $X_{vib}(t_{\infty})$  is denoted $X_{vib}(\infty)$, where 
\begin{equation} \label{eq3}
X_{vib}(\infty)=\frac{6 k_{B}T}{M} \frac{1}{3N-3}\sum_{\lambda} \frac{1}{\omega_{\lambda}^2}.
\end{equation}
When evaluated numerically, vibrational time correlation functions develop long-time recursive noise, a finite-$N$ effect. We eliminate this noise by setting $X_{vib}(t)$ to $X_{vib}(\infty)$ at $t>t_{\infty}$. This truncation can be seen in figure \ref{fig2}.

V-T theory is expressed in equations for the three time intervals of figure \ref{fig2}, as developed in \cite{MSD1}. 

\noindent Vibrational interval:  \qquad \qquad \qquad $ 0 \leq  t \leq \tau_{D}$    
\begin{equation} \label{eq4}
      \qquad  X_{VT}(t) =X_{vib}(t);  \\
\end{equation}
Crossover interval:  \qquad \qquad  \qquad $\tau_{D} \leq t \leq \tau_{RW} $
\begin{equation} \label{eq5}
      \qquad X_{VT}(t) =X_{vib}(t) +  \nu (t-\tau_{D}) S^2;\\
\end{equation}
Steady State transit random walk: \qquad $t \geq \tau_{RW}$                     
\begin{equation} \label{eq6}
     \qquad  X_{VT}(t) =X_{VT}(\tau_{RW}) +  \nu (t-\tau_{RW}) (\delta R)^2.
\end{equation}

In equation~(\ref{eq6}), $X_{VT}(\tau_{RW})$ is equation~(\ref{eq5}) evaluated at $t=\tau_{RW}$. The delay time $\tau_{D}$ is the end of the vibrational interval,  $\tau_{RW}$ is the beginning of the random walk interval, $\nu$ is the single-atom transit rate, $S$ is the step length of the crossover walk, and $\delta R$ is the step length of the random walk. We shall analyze figure \ref{fig2} with the equations (\ref{eq4})-(\ref{eq6}), and will calibrate the parameters from $X_{MD}(t)$ and $X_{vib}(t)$. The results are listed in table I, along with the calibrations at 395.1K for comparison.

\begin{figure} [h]
\includegraphics[height=5.0in,width=5.0in,keepaspectratio]{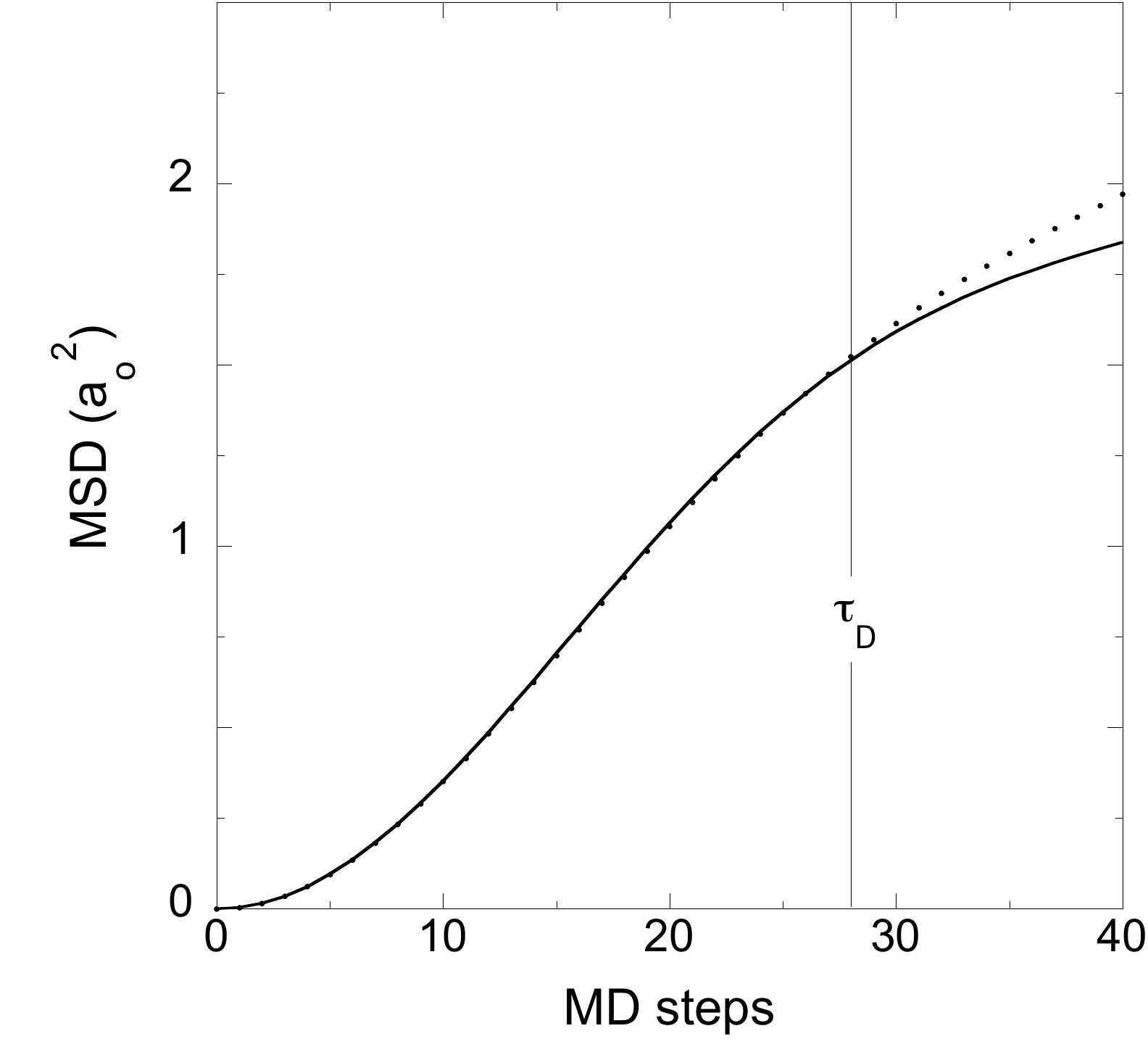}
\caption{At 204.6K: Dots are $X_{MD}(t)$ and  line is $X_{vib}(t)$. Dots are at every $\delta t$. }
\label{fig3}
\end{figure}

Transits have been observed as the correlated motion of small local groups of atoms in equilibrium MD trajectories at low $T$ \cite{WCC2001}. For the time correlation functions we study, only the separate one-atom transit is required. The transit motion can be projected out of a single atom trajectory, leaving that trajectory as vibrational motion about a fixed equilibrium position, plus the transit-induced motion of the atomic equilibrium position. The latter, heavily averaged over transits, is called the ``transit motion". In the equilibrium liquid, the system-wide transit motion consists of a steady random walk of each atom, with step rate $\nu$ and step distance $\delta R$. For a time correlation function started at $t=0$, we identify $\tau_{D}$ as the time required to resolve and begin to measure the ongoing transits. Some time after $\tau_{D}$, the MD system is fully measuring every transit as it is completed. Additional details of transit theory are discussed in section 3 of \cite{SISF2}.

$\tau_{RW}$ is defined as the time when $X_{MD}(t)$ begins its ultimate pure diffusive behavior (section 3 of  \cite{MSD1}). This requires two conditions: $X_{MD}(t)$ must be fully measuring the steady transit random walk, and $X_{vib}(t)$ must have reached its ultimate constant value. At 395.1K, $\tau_{RW}$ is determined directly from the graph of $X_{MD}(t)$, automatically satisfying both conditions (see figure 3 of \cite{MSD1}). At 204.6K, this procedure provides poor resolution. This would not be a problem, since it merely leads to a wide range of acceptable calibration. However, theory alone will give us a more informative calibration. From the 395.1K calibration, we know the first condition is satisfied before $60\delta t$. At 204.6K, the transit rate is so low that transit damping of $X_{vib}(t)$ is ineffective, leaving $X_{vib}(t)$ fully present up to $t_{\infty}$, as shown in figure \ref{fig2}. But since we set $X_{vib}(t)=X_{vib}(\infty)$ for $t\geq t_{\infty}$, to get rid of the long-time vibrational noise, the second condition is satisfied at $\tau_{RW}= t_{\infty}$. Moreover, this result holds at all lower $T$, since the transit rate only decreases with decreasing $T$, so that $\tau_{RW}= t_{\infty}$ at $T \leq$ 204.6K.

Equation~(\ref{eq4}) tells us that the theoretical  $X_{VT}(t)$ is given by its vibrational contribution alone in the vibrational interval. Figure \ref{fig3} confirms equation~(\ref{eq4}) to high accuracy, and also confirms that the time  $\tau_{D}$ when $X_{MD}(t)$ moves away from $X_{vib}(t)$ is precisely the same here as at 395.1K (see table I).

\begin{figure} [h]
\includegraphics[height=5.0in,width=5.0in,keepaspectratio]{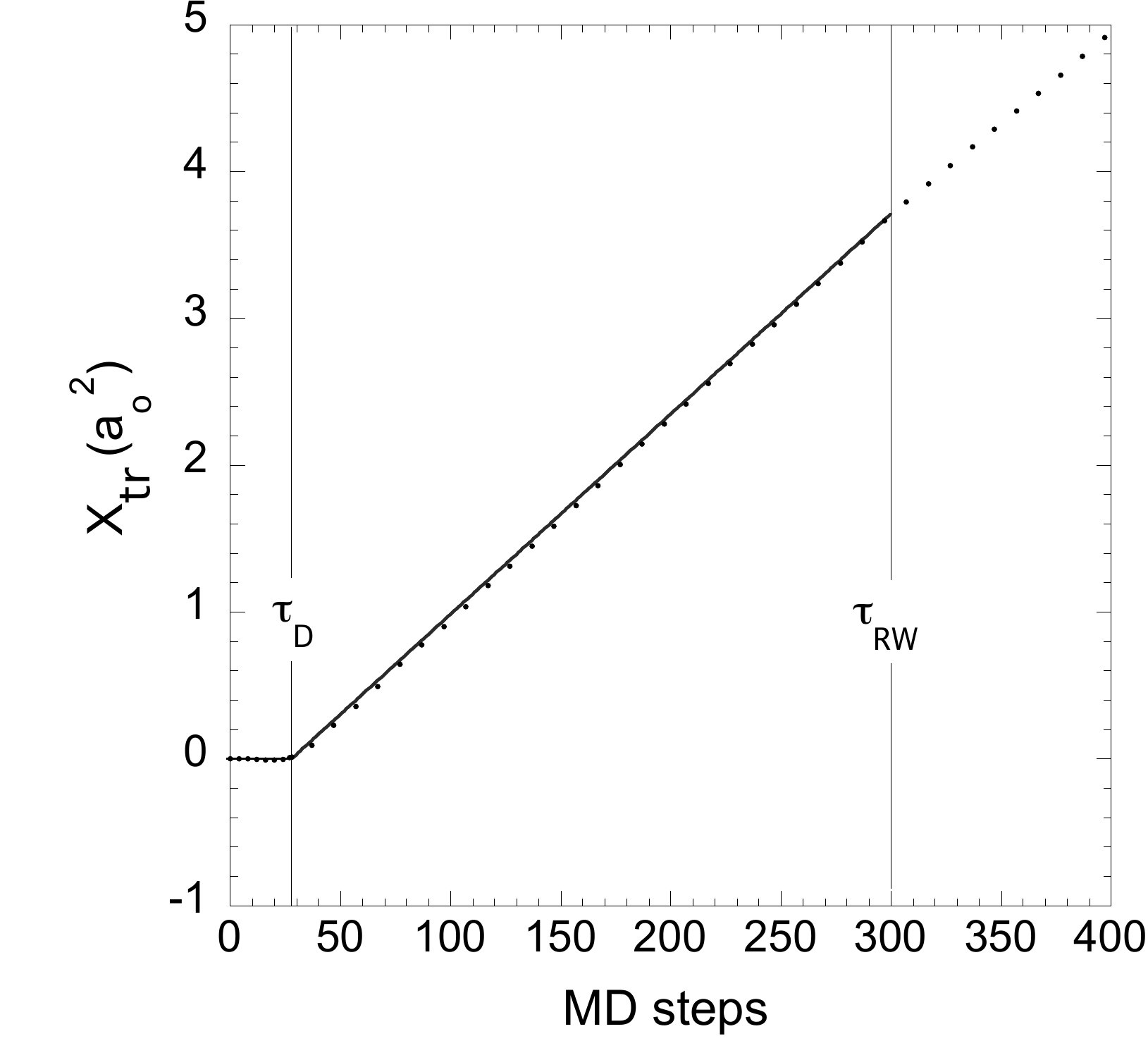}
\caption{At 204.6K: Dots are $X_{MD}(t)-X_{vib}(t)$ and  straight line  segments are the approximation for $X_{tr}(t)$. Dot timing as in figure 2. }
\label{fig4}
\end{figure}

Equation~ (\ref{eq5}) is the crossover theory developed at 395.1K. The last term in equation~(\ref{eq5}) is $X_{tr}(t)$ and represents a random walk which we call the ``crossover walk".  The crossover walk is linear in $t$, and so is expressed in figure \ref{fig4} as a straight line fitting $X_{MD}(t) - X_{vib}(t)$ at $\tau_{D}$ and $\tau_{RW}$. This approximation for  $X_{tr}(t)$ was chosen for its tractability and simple physical meaning \cite{MSD1}. The overall accuracy of the approximation is excellent in figure \ref{fig4}, as it is at 395.1K. The calibration finds $S$ close to $\delta R$, table I. We shall rationalize this result in section IV.

The last term in equation~(\ref{eq6}) is  $X_{tr}(t)$ in the form of the steady transit random walk; it constitutes the entire time dependence in equation (\ref{eq6})  and continues indefinitely.  The steady random walk is calibrated from the measurement of  the self diffusion coefficient $D$: a straight line is fitted to a long equilibrium $X_{MD}(t)$ segment, omitting data at $t\leq\tau_{RW}$, and the straight line slope is 
\begin{equation} \label{eq8}
\dot{X}_{MD}=6D=\nu (\delta R)^{2}.
\end{equation} 
The first equality is hydrodynamics and the second is the Einstein random walk. $\delta R$ is considered independent of $T$, so its calibration in table I is the same at 204.6 and 395.1K. The transit rate $\nu$ is then determined from equation (\ref{eq8}). 

\begin{figure} [h]
\includegraphics[height=5.0in,width=5.0in,keepaspectratio] {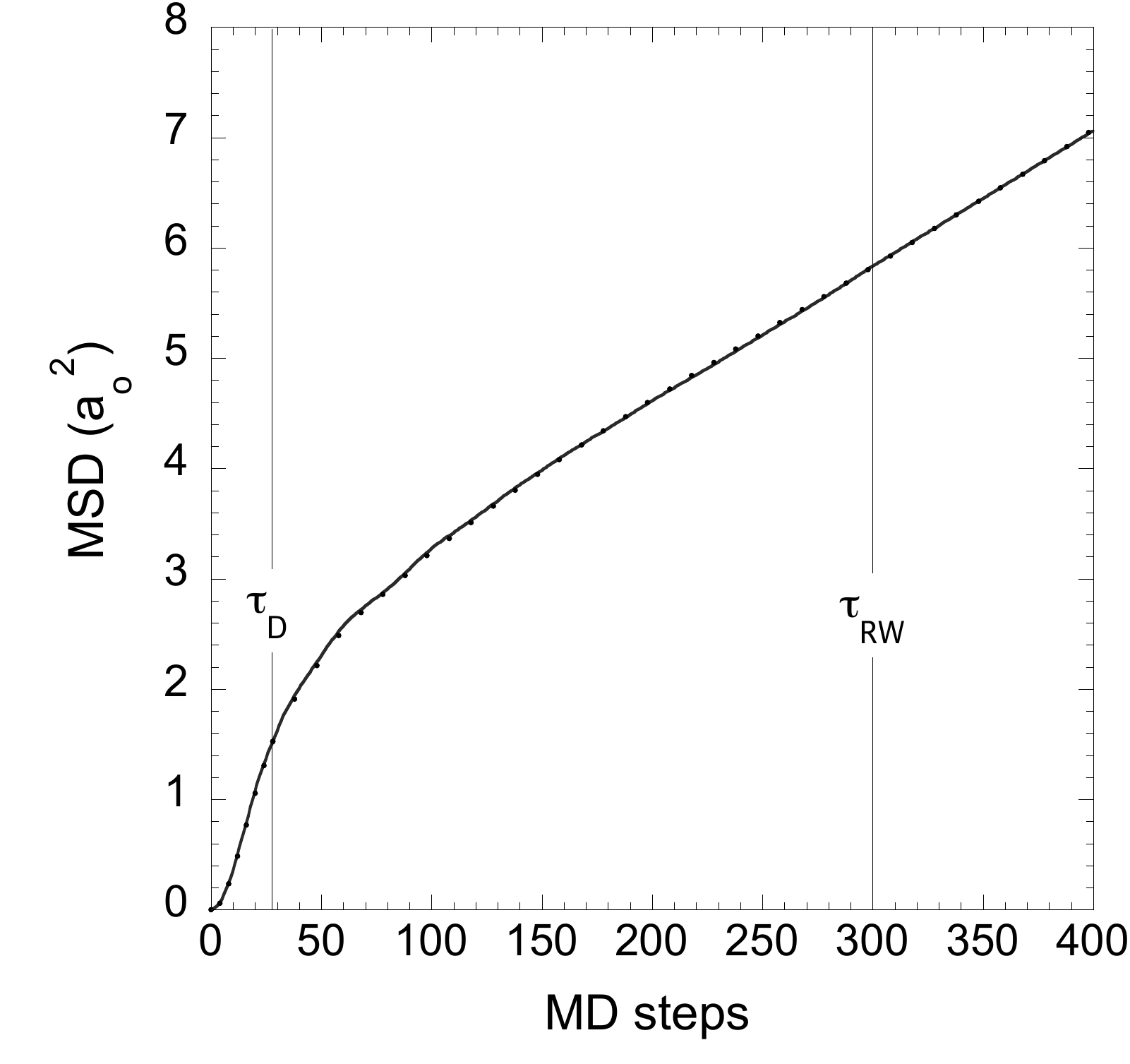}
\caption{At 204.6K: Dots are $X_{MD}(t)$ and  line is $X_{VT}(t)$.  Dot timing as in figure 2. }
\label{fig5}
\end{figure}

Figure \ref{fig5} shows excellent agreement of $X_{VT}(t)$ with $X_{MD}(t)$ for all $t$ to beyond $\tau_{RW}$. The maximum relative error in $X_{VT}(t)$ is 2.4$\%$ near 120$\delta t$, reproducing almost exactly the maximum error of 2.5$\%$  at 395.1K.  
From equations (\ref{eq5}) and (\ref{eq6}), $X_{VT}(t)$  has zero error at $\tau_{RW}$, and will continue with zero error  because it has the straight line slope of $X_{MD}(t)$.

\begin{table}[h] 
\caption{\label{table1}Calibration of the motional parameters at study temperatures. Rows 1 and 2 are calibrations from $X_{MD}(t)$ (dots in figure \ref{fig6}), and rows 3-6 are from theory (line in figure \ref{fig6}). $T_{m} = 371.0$K for Na, and  the MD time step is $\delta t = 7.00288$~fs.}
\begin{ruledtabular}
\begin{tabular}{ccccccc}
$T$(K)&$\nu$ (ps$^{-1}$)&$S$(a$_{0})$&$\delta R$(a$_{0})$&$\tau_{D}(\delta t)$&$\tau_{RW}(\delta t)$\\
\hline
395.1&3.9&1.46&1.75&28&60\\
204.6&0.573 &1.84&1.75&28&300\\
\hline
204.6&0.559 &1.75&1.75&28&300\\
128.7&0.0514&1.75&1.75&28&300\\
94.5&0.00500&1.75&1.75&28&300\\
74.0&0.000443&1.75&1.75&28&300\\
\end {tabular}
\end{ruledtabular}
\label{Table1}
\end{table}

\section{Dynamic slowing down}

As we shall see, dynamic slowing down is just beginning to appear in liquid Na at 204.6K. We shall have to work at significantly lower temperatures to study this behavior.  Moreover, we cannot use MD for this work, because our MD system will not remain an equilibrium liquid to sufficiently low $T$.  We therefore set out to employ V-T theory to make the study. The required theoretical information is (a) the apriori derived and calibrated vibrational contribution, equations~(\ref{eq2}) and (\ref{eq3});  (b) the liquid time-evolution equations~(\ref{eq4})-(\ref{eq6}); and (c) the five motion parameters calibrated for liquid Na in the ``low-$T$ regime", below 204.6K. All is ready except for item (c); this can be accomplished from information determined at $T\geq$ 204.6K.

Referring to table I, $\tau_{D}$ is independent of $T$ from 395.1 to 204.6K, and we shall allow it to remain so. A comment on the physical nature of $\tau_{D}$ is given in section V.A.  In section III, $\tau_{RW}$ is set to $t_{\infty}$ at 204.6K and all lower $T$. The reason why $S$ approaches $\delta R$ as $T$ decreases to 204.6K is as follows. Some transit motion is lost (never measured) in the vibrational interval, as mentioned following equation (8) in \cite{MSD1}. The amount of transit motion lost must be proportional to $\nu(T)$, hence the lost information decreases as $T$ decreases, making $S$ approach $\delta R$. Within calibration errors, $S$ and $\delta R$  are already equal at 204.6K, table I, so we set $S=\delta R$  at all $T \leq$ 204.6K.

\begin{figure} [h]
\includegraphics[height=5.0in,width=5.0in,keepaspectratio]{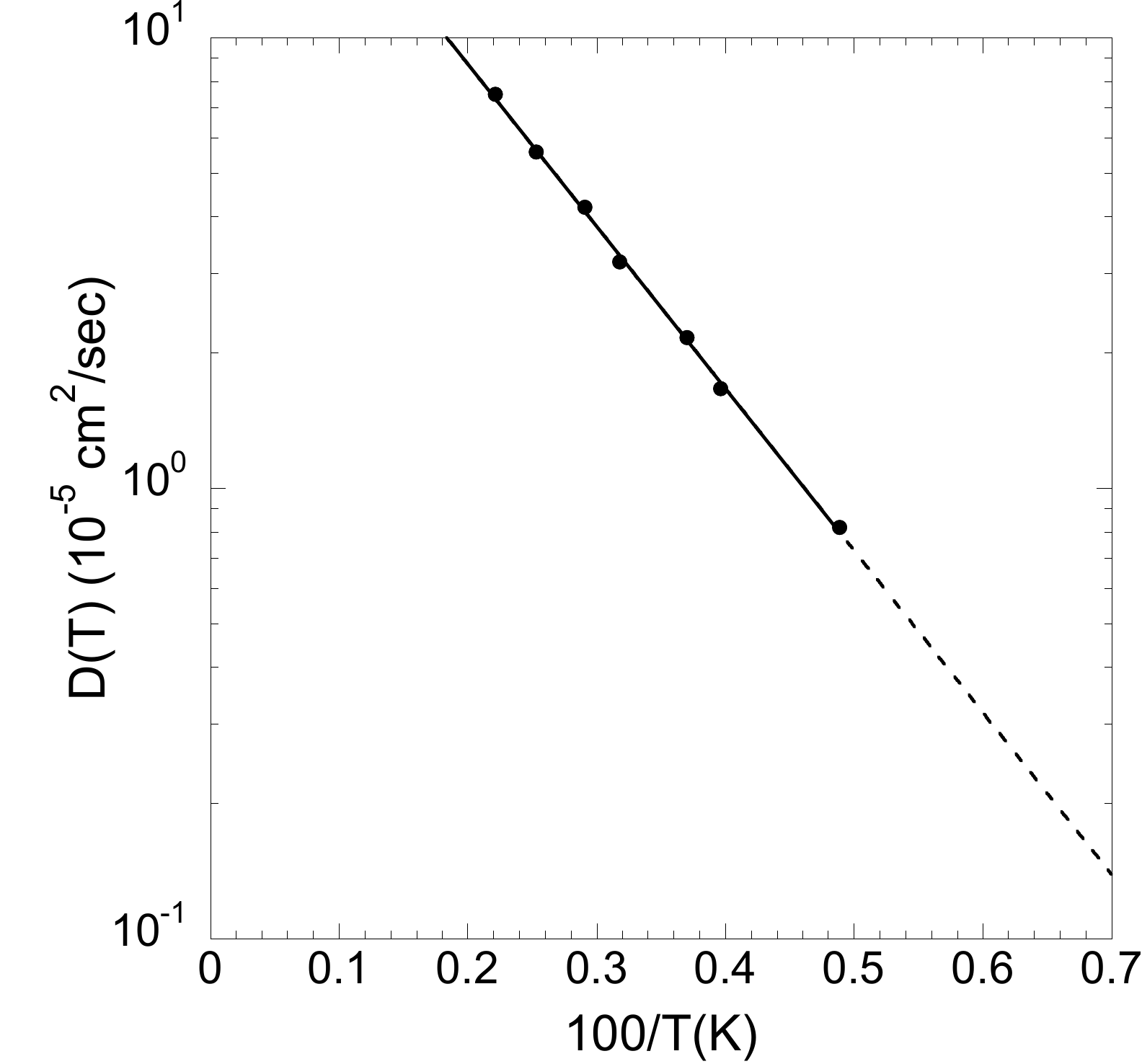}
\caption{Arrhenius graph for liquid Na. Dots are $D_{MD}(T)$ at $T$ down to 204.6K. Straight line is fitted to the $\log D_{MD}(T)$  points (solid line), and is extrapolated for the low-$T$ theory below 204.6K (dashed line).}
\label{fig6}
\end{figure}

The last needed parameter for the low-$T$ regime is $\nu(T)$, and is to be calibrated from $D_{MD}(T)$, according to equation (\ref{eq8}).  In figure~\ref{fig6}, our $D_{MD}(T)$ points show  Arrhenius behavior and are fitted to a straight line in $T^{-1}$:
\begin{equation} \label{eq7}
D_{MD}(T)=Ae^{-\beta\phi},
\end{equation}
where $A=45.78(10^{-5}$cm$^2$/s), $\beta = 1/k_{B}T$ and $\phi = 828.1$K. The transit rate is quite low at 204.6K, and we presume the Arrhenius behavior continues with decreasing $T$.
Equation (\ref{eq7}) is extrapolated, as indicated in figure \ref{fig6}, and the corresponding liquid-phase $\nu(T)$ values are listed in table I.

\begin{figure} [b]
\includegraphics[height=5.0in,width=5.0in,keepaspectratio]{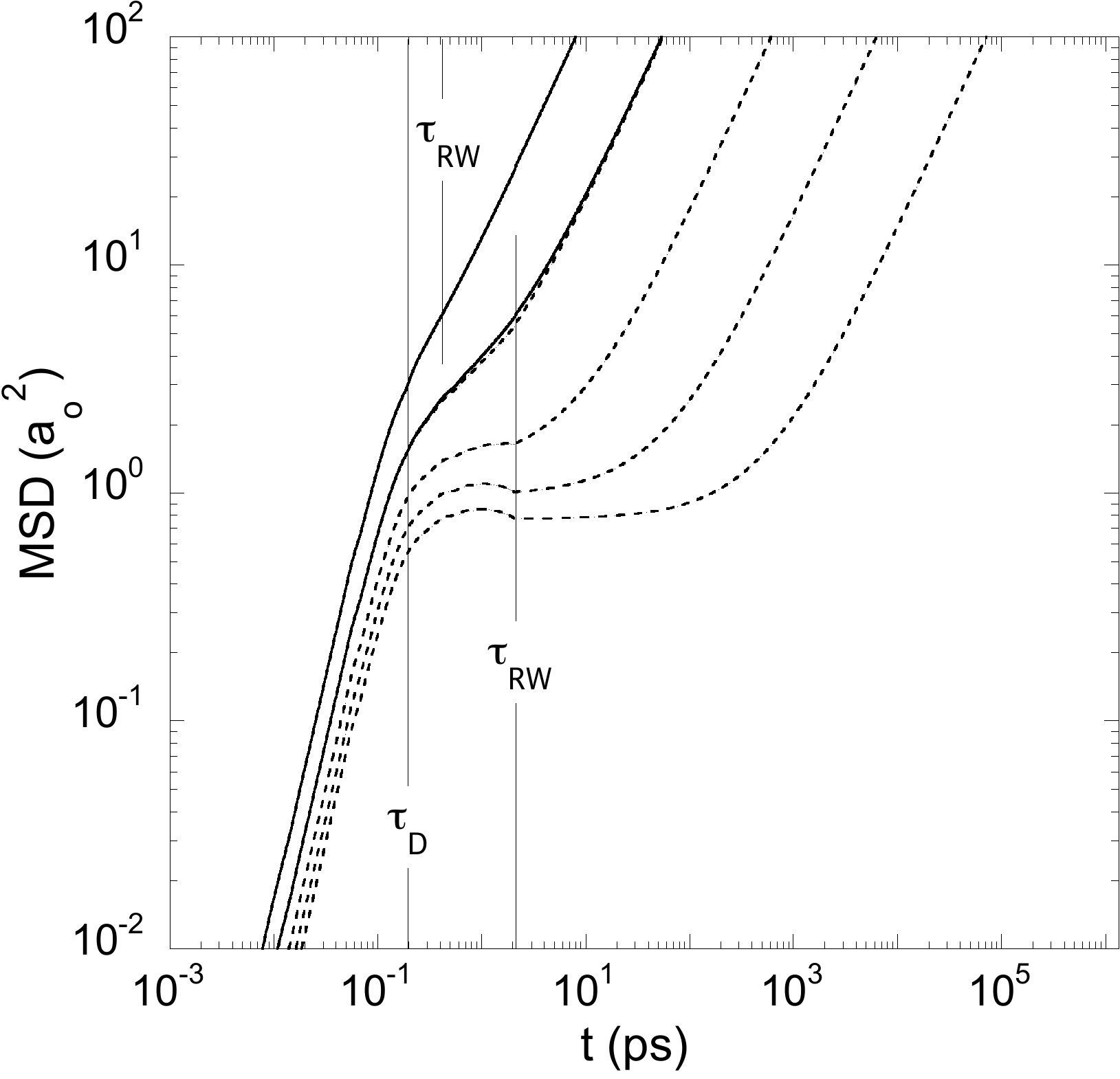}
\caption{All curves are the theoretical  $X_{VT}(t)$, from the top at $T=395.1, 204.6, 204.6, 128.7,94.5, 74.0$K. The two highest curves (solid) are $X_{VT}(t)$ calibrated from $X_{MD}(t)$, and the four  lowest curves (dashed) are the low-$T$ approximation for $X_{VT}(t)$. Difference between the two curves at 204.6K is  visible around 1ps.}
\label{fig7}
\end{figure}

Figure~\ref{fig7} shows the K-A graph of the theoretical $X_{VT}(t)$ for liquid Na.  The solid-line curves at the two highest temperatures are fitted to MD data, hence are near-exact copies of $X_{MD}(t)$. The dashed-line curves  at $204.6$K and below are the low-$T$ approximation. This approximation agrees closely with the complete V-T theory at $204.6$K. 

In figure \ref{fig7}, the appearance and extension of the plateau with decreasing $T$ expresses dynamic slowing down. This is clearly a two-step process, a precursor on the crossover, followed by the plateau, expressed respectively by the $X_{tr}(t)$ contributions in equations~(\ref{eq5}) and (\ref{eq6}). The bump ahead of the plateau is the temporary rise of $X_{vib}(t)$  above $X_{vib}(\infty)$, apparent in figure~\ref{fig2},  and the kink at $\tau_{RW}$ is the slope discontinuity in figure \ref{fig2}. The same precursor appears as a dip, i.e. with negative sign, in $F_{vib}^{s}(q,t)$, where the feature is referred to as the vibrational excess (figures~1 and 4 of \cite{SISF1}). The vibrational excess also appears in MD data. It is barely visible in the BMLJ liquid (figure~1 of \cite{KA1994}; figure~2 of \cite{KA1995a}; figure~2 of \cite{KA1995b}).  However, the vibrational excess  is quite pronounced in silica (figure~2(a) of \cite{HKB1998}; figures~3b and 5b  of \cite{Kob1999}). Figure \ref{fig7}  is overall quite similar to these K-A graphs from MD. We shall discuss this similarity in terms of the underlying atomic motion in section V.B.

The comparison of liquid Na with well-known glass forming liquids is more sensible than one might suppose. The reason is that all comparisons are made for the liquid phase, which excludes the glass transition. Kob and Andersen stress that critical behavior occurs in the equilibrium liquid at $T$ above the MCT critical temperature $T_{C}$ \cite{KA1994, KA1995a}. For this reason those authors work at $T$ from 5.0 to 0.466, staying well above $T_{C}=0.435$ for their BMLJ system. They also verify their system is not undergoing a glass transition (figure 1 of \cite{KA1995a}). In the present liquid Na study, we endow the system with the same two properties by having developed and calibrated the theory so as to describe equilibrium atomic motion within the random valley distribution.

\section{Discussion and conclusions}

\subsection{Liquid vibrational theory}

The observation that $\tau_{D}$ is independent of $T$, table I, is a strong indication that $\tau_{D}$  is of vibrational origin. A useful vibrational time is the mean vibrational period, defined as $\tau_{vib}=2\pi/\sqrt{\left<\omega ^{2}\right>}$, where the average is over the vibrational frequencies. Pictorially, $\tau_{vib}$ is the time for an atom to move a circumferential path length on its vibrational surface. For the present Na system, $\tau_{vib}\approx 59.0 \delta t$ (see equation (15)  and table II of \cite{VACF1998}), and from table I this implies  $\tau_{D}\approx 0.5 \tau_{vib}$. We are currently developing a transit geometry approximation that will rationalize this result. 

The vibrational formulation we have constructed has a set of control factors that make it useful in condensed matter theory. First, the V-T decomposition of the motion is defined: The 3$N$-dimensional potential surface rising from a structure is harmonically extended to infinity, so that the vibrational Hamiltonian and its eigenstates are defined. Then the intervalley intersections lost in the extension are made part of the transit problem. 
Second, it is the only such theory capable of producing exact formulas for statistical mechanical functions. In this application, the theory is significantly simplified by requiring only a single random valley. Finally, since it is fully prescribed, with no adjustable parameters, it is the same uniform theory in all applications for all materials. 

In the vibrational interval, $0 \leq t \leq \tau_{D}$, we have a set of ten independent test calculations that yield precisely the same result. The tests are comparisons of the vibrational contribution alone with MD data for $X_{MD}(t)$   at 395.1 and 204.6K, and with MD data for  $F_{MD}^{s}(q,t)$ at 395.1K for eight $q$ \cite{MSD1, SISF2}.The result is that the vibrational contribution is in highly accurate agreement with the MD data in every case. The particular comparison made here is shown in figure 3. This result, along with the analytic properties of the vibrational theory mentioned above, presents a robust foundation in the many-body formulation of time correlation functions.

According to equation (\ref{eq3}),  $X_{vib}(t)$  is proportional to $T$ (in the present classical statistical mechanics). This factor comes from the canonical weight $e^{-\beta H_{vib}}$, where $\beta = 1/k_{B}T$ and  $H_{vib}$ is the vibrational Hamiltonian. Now since $X_{vib}(t)=X_{MD}(t)$ in the vibrational interval, then $X_{MD}(t)$ must also be proportional to $T$ to very high accuracy in $0 \leq t \leq \tau_{D}$. The theory has uncovered a scaling property of the MD data. We suppose this scaling in $T$ will be generally found among monatomic liquids, and some more complicated liquids as well. Incidentally, the $T$ scaling does not hold in $F_{vib}^{s}(q,t)$,  or holds only to an approximation, because the $T$ factor is inside the fluctuation (equations (4)-(6) in \cite{SISF2}).

\subsection{Liquid mean square displacement theory}

In their testing of mode coupling theory, Kob and Andersen cite the key role of  $D(T)$ in critical dynamics \cite{KA1994, KA1995a}. For liquid BMLJ, $D(T)$ follows the critical power law in $T$, and transfers that behavior to various time correlation functions. Kob and Andersen address the pure diffusive process, which operates from the start of the plateau, and for which $X_{MD}(t)$ has the time dependence $6D(T)t$. The same diffusive process operates in liquid silica, which, in contrast, follows Arrhenius dynamics at temperatures where dynamic slowing down takes place (figures 3(b) and 4(b) of \cite{ Kob1999}). In both liquids, BMLJ and silica, the plateau extends without limit as $D(T)$ approaches zero. The difference is that $D(T) \rightarrow 0$ as $T\rightarrow T_{C}$ in critical dynamics, and as $T\rightarrow 0$ in Arrhenius dynamics.

We set out in this research to test the ability of V-T theory to explain the $T$ dependence of $X_{MD}(t)$ for $T\leq$ 395.1K. $T$ dependence of $X_{vib}(t)$ is prescribed in equations (\ref{eq2}) and (\ref{eq3}), and $T$ dependence of $X_{tr}(t)$ is contained in equations (\ref{eq5}) and (\ref{eq6}), with parameters listed in table I. As $T$ decreases to 204.6K, the dynamic slowing down precursor develops in the crossover and ``freezes" into place (section III; also table I). The precursor is visible in figure~\ref{fig7} at $T\leq$ 128.7K. Below 128.7K, the precursor and plateau decrease in magnitude as $T$ decreases, a vibrational effect, and the plateau extends in time as $T$ decreases, a transit effect. This transit effect is precisely the pure diffusive contribution addressed by the slowing dynamics studies cited above. Hence the diffusive $t$ dependence $6D(T)t$ is contained in equation (\ref{eq6}), with assistance from equation (\ref{eq8}). Moreover, the starting time $\tau_{RW}$ of the pure diffusive motion, and the magnitude of $X_{VT}(\tau_{RW})$, are given by the time evolution up to  $\tau_{RW}$, and are also contained in equation (\ref{eq6}).

Ultimately, the systems we have discussed remain ergodic. Critical behavior ends before $T_{C}$ is reached for BMLJ  (figure 5.19  of \cite{BK2005}), while liquid silica joins Na with ergodic Arrhenius dynamics (figure 5.20 of \cite{BK2005}). 

Figures and discussions throughout this manuscript confirm that V-T theory does explain the widely varying $T$ dependences appearing in figure \ref{fig7}. The present work extends V-T theory from space-time dimensions into the dimension $T$.  The logical next research topic will be to apply the present analysis at $T$ increasing from 395.1K. This will present new and interesting challenges.

\subsection{Comment on symmetric states}

In the glass transition the system moves among symmetric valleys, and the slowing dynamics is more complicated than for the liquid. Berthier and Biroli observe that ``slow dynamics in the supercooled state is due to the existence of some local order reminiscent of the crystal structure" (section VI.C of \cite{BB2011}). Such order identifies the symmetric structures, defined in section II, from \cite{DCW1997}. For Na, we have shown that the symmetric structures are clearly distinguished from random structures by symmetry measures of their Voronoi neighbors \cite{CW1999}. The appearence and growth of local symmetry  in supercooled LJ binary mixtures \cite{PSDH2010,DCGP2007} and colloidal systems \cite{TKSW2010} is currently under study.

\acknowledgements{We are pleased to thank  A Voter, C Reichardt and B Clements for helpful and encouraging discussions. This research is supported by the Department of Energy under Contract No. DE-AC52-06NA25396.}

\bibliography{TheoryforFd} 

\end{document}